\documentclass[prc,showpacs]{revtex4}
\usepackage{graphicx}
\usepackage{dcolumn}
\usepackage{bm}

\begin{document}

\title{Shape transitions in neutron-rich Yb, Hf, W, Os, and Pt isotopes
within a Skyrme Hartree-Fock + BCS approach}

\author{P. Sarriguren}
\affiliation{Instituto de Estructura de la Materia, CSIC, Serrano
123, E-28006 Madrid, Spain}

\author{R. Rodr\'{\i}guez-Guzm\'an}
\affiliation{CEA-Saclay DSM/IRFU/SPhN, F-91191 Gif-sur-Yvette, France}

\author{L.M. Robledo}

\affiliation{Departamento  de F\'{\i}sica Te\'orica C-XI,
Universidad Aut\'onoma de Madrid, 28049-Madrid, Spain}

%\email{sarriguren@iem.cfmac.csic.es}

\date{\today}

\begin{abstract} 
Self-consistent axially symmetric Skyrme Hartree-Fock plus BCS calculations
are performed to study the evolution of shapes with the number of nucleons 
in various chains of Yb, Hf, W, Os, and Pt isotopes from neutron number 
$N$=110 up to $N$=122. Potential energy curves are analyzed in a search 
for signatures of oblate-prolate phase shape transitions and results from 
various Skyrme and pairing forces are considered. Comparisons with results
obtained  with the Gogny interaction as well as with relativistic mean field 
calculations are  presented. The role of the $\gamma$ degree of freedom
is also discussed. 

\end{abstract}

\pacs{21.60.Jz, 27.70.+q, 27.80.+w}

\maketitle

\section{Introduction}
\label{Intro}
 
The study of the equilibrium shapes that characterize the ground states
of atomic nuclei, as well as the transitional regions where shape changes
occur, have been the subject of a large number of theoretical and experimental 
works (for a review, see, for example, Ref. \cite{review} and references 
therein). In particular, the complex interplay between different competing 
degrees of freedom, taking place in transitional nuclei, offers the possibility 
to test microscopic descriptions of atomic nuclei under a wide variety of 
conditions. In this context, mean field approximations based on effective 
interactions with predictive power all over the nuclear chart, which are a 
cornerstone to almost all microscopic approximations to the nuclear many-body 
problem, appear as a first theoretical tool to rely on when looking for 
fingerprints on nuclear phase shape transitions.

Nowadays, systematic mean field studies have become possible due to two main 
reasons. First, important advances have already been made in the fitting 
protocols providing effective nucleon-nucleon interactions with predictive 
power all over the nuclear chart. Popular energy density functionals for 
calculations along these lines are the nonrelativistic Gogny \cite{gogny} 
and Skyrme \cite{vautherin,Bender-Review} ones, as well as different 
parametrizations of the relativistic mean field Lagrangian 
\cite{Bender-Review,lala-ref}. Second, it has also become possible to recast 
mean field equations in terms of efficient minimizations like successive 
iteration methods or the so called gradient method (see, for example, Refs. 
\cite{Bender-Review} and \cite{gradient-egido}). The last point is very 
important when constrained calculations are performed since there, in addition 
to the usual constraints on both neutron and proton numbers, other external 
fields could also be added. A typical situation is that where both $\beta$ and 
$\gamma$ degrees of freedom are taken into account. In addition, mean field 
approximations are based on product trial wave functions, used to minimize 
a given energy density functional. Such products break several symmetries of 
the underlying nuclear hamiltonian, allowing the use of an enlarged 
Hilbert space. Within this space, static correlations associated with collective 
modes (e.g., deformation) are incorporated at the cost of a moderate effort. 
These are the main reasons why the mean field framework can be considered as 
a valuable starting point for microscopic nuclear structure studies.

%%%%%%%%%%%%%%%%%%%%%%%%%%%%Fig1%%%%%%%%%%%%%%%%%%%%%%%%%%%%%%%%%%%%%%%%%%%%%%%%%%%%%%%%
\begin{figure*}[b]
\centering
\includegraphics[width=110mm]{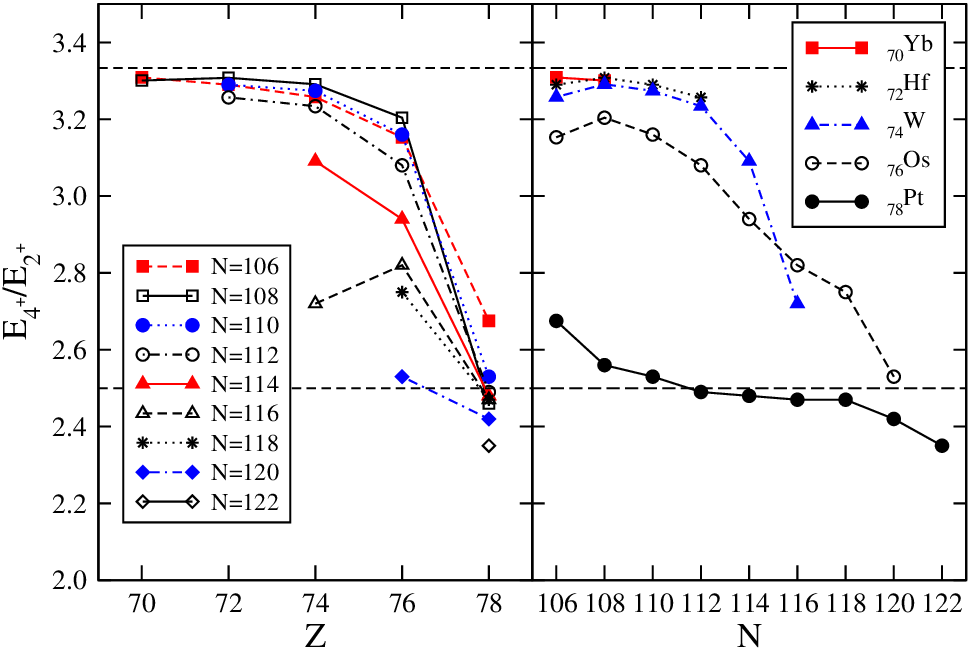}
\caption{(Color online) Experimental ratio of the first $4^+$ and $2^+$ energies 
for various isotone (left) and isotope (right) chains. Dashed horizontal lines 
are plotted at 3.33 and 2.5. For details, see the main text.}
\label{fige42}
\end{figure*}
%%%%%%%%%%%%%%%%%%%%%%%%%%%%%%%%%%%%%%%%%%%%%%%%%%%%%%%%%%%%%%%%%%%%%%%%%%%%%%%%%%%%%%%%%
 
In this paper we plan to use this approach to get insight into nuclear shape 
transitions around $^{190}$W. This mass region is particularly interesting 
because it lies below doubly magic numbers and small islands of oblate 
deformations might be favored energetically. It is also characterized by 
the strong competition between oblate and prolate configurations (i.e., 
shape coexistence) and special interest has received the case of $^{190}$W 
\cite{podolyak,caamano,walker2006}. So far, much data have been collected, 
especially on the energy spacing of the lowest-lying states in even-even 
systems with  mass number $A$=170-200. 
Spectroscopic studies on these nuclei have become possible by exploiting the
decay of K-isomers, which is also a well known feature characterizing this 
mass region \cite{nature}.
This turns into a significant amount of information on the global behavior of 
these nuclei. For example, the excitation energy of the first $2^+$ state is 
used to correlate the extent of quadrupole deformation, and the ratio of the 
first $4^+$ to the first $2^+$ $(E_{4^+}/E_{2^+})$ can be used, in simple 
models, to distinguish between an axially symmetric deformed rotor 
$(E_{4^+}/E_{2^+}=3.33)$, a spherical vibrational nucleus 
$(E_{4^+}/E_{2^+}=2.0)$ and a triaxial rotor 
$(E_{4^+}/E_{2^+}=2.5)$ \cite{casten2000,jolie}. 

The experimental situation \cite{firestone,podolyak,caamano} concerning the 
ratio $E_{4^+}/E_{2^+}$ for even-even isotopes in this region is summarized 
in Fig. \ref{fige42}. Clear signatures of shape transitions, which are the 
main objective of this paper, are visible from this figure. In the left panel, 
we have plotted the ratio $E_{4^+}/E_{2^+}$ as a function of $Z$ for nine 
isotone chains from $N$=106 up to $N$=122. As can be observed, there is a clear 
tendency toward the prolate rotational limit ($E_{4^+}/E_{2^+}=3.33$) in the 
isotone chains below $N$=116 as the number of protons decreases. In the case 
$N$=116 a change of tendency is also observed between $Z$=76$\,(^{192}$Os) 
and $Z$=74$\, (^{190}$W) that has been interpreted as a subshell effect \cite{mach}. 
On the other hand, the heavier $N$=116 isotones $^{192}$Os and $^{194}$Pt, are 
well known examples of $\gamma$-soft nuclei \cite{wu96} and then, it is natural 
that the $\gamma$ degree of freedom might also play an important role in $^{190}$W, 
as it can be deduced from the close value of the $E_{4^+}/E_{2^+}$ ratio to the 
2.5 limit. In the right panel of Fig. \ref{fige42}, the ratio $E_{4^+}/E_{2^+}$ 
has been plotted for various isotopes as a function of the neutron number. As 
can be seen, the lighter Yb, Hf, W, and Os isotopes have a ratio which is very 
close to the rotational limit. On the other hand, for heavier isotopes the ratio 
decreases up to values close to 2.5 and the tendency is to continue toward the 
spherical limit. Finally, the  Pt isotopes have ratios that indicate a 
pronounced $\gamma$-soft character.

From the theoretical point of view, a shape transition from prolate to oblate 
shapes, as the number of neutrons is increased in this mass region, has 
already been predicted within different models 
\cite{wu96,naza90,wheldon2000,naik,fossion,stevenson05}. For instance, a 
pioneer study within the shell correction method with a deformed Woods-Saxon 
potential and monopole pairing interaction has been used \cite{naza90} to 
discuss the prolate-oblate shape change in Os isotopes. It has been found 
that the shape change takes place between $N$=114 and $N$=116, while similar 
studies \cite{wheldon2000}, based on total routhian surface calculations, also
suggested this transition to occur at $N$=118. On the other hand, various 
collective models were tested in Ref. \cite{wu96} to describe the $E2$
properties of the low-lying states in several rare-earth Os and Pt isotopes.
The results indicated that the data were consistent with the description
of these nuclei as being $\gamma$-soft. Shape predictions from relativistic 
and nonrelativistic studies with angular momentum projection for Hf, W, and 
Os isotopes were also compared in Ref. \cite{naik}. While relativistic 
calculations predicted a majority of prolate shapes, cycling changes in the 
sign of the quadrupole parameters were observed in the case of nonrelativistic 
calculations. More recently, the shape transition in both Os and Pt isotopes 
has been studied \cite{fossion} within the relativistic mean-field (RMF) 
approximation using the parametrization NL3, perhaps the best parameter set
ever fitted for the RMF Lagrangian, together with a finite range Brink-Boeker  
pairing interaction and also within a nonrelativistic self-consistent axially 
deformed Hartree-Fock (HF) framework based on a separable monopole interaction 
\cite{stevenson05,stevenson01}. 

Nevertheless, and to the best of our knowledge, no systematic study of shape 
transitions in this mass region has been carried out yet within the framework 
of Skyrme Hartree-Fock calculations with pairing correlations, which is at 
present one of the state of the art mean field descriptions (see, for example, 
\cite{vautherin,Bender-Review,ev8}). Therefore, in this work we study the 
ground state shape evolution in five isotopic chains, namely, Yb, Hf, W, Os, 
and Pt from $N$=110 up to $N$=122 in an attempt to get first hints on nuclear 
shape transitions for these nuclei. In particular, we will first keep axial 
symmetry as a self-consistent symmetry \cite{rs} and construct 
potential energy curves (PECs) as functions of the (axially symmetric) 
mass quadrupole moment for the above mentioned chains of nuclei. These PECs 
are obtained from the corresponding constrained Skyrme Hartree-Fock + BCS 
calculations, using the forces Sk3 \cite{sk3}, SLy4 \cite{sly4}, and 
SLy6 \cite{sly4} in the particle-hole channel and different recipes for 
pairing correlations. The axially symmetric calculations should be regarded 
as a first step and will allow us to disentangle the sensitivity of our 
predictions with respect to the method employed to solve the deformed
HF+BCS equations (i.e., discretization in a cartesian mesh or expansion
into an axially symmetric harmonic oscillator basis), as well as to the effective 
Skyrme and pairing interactions. Additionally, we will also take the chance for 
a comparison with results obtained with the parametrization D1S \cite{d1s} of 
the Gogny interaction \cite{gogny}. Let us remark that the main intention in 
this work is to obtain first hints on shape transitions around the nucleus 
$^{190}$W using effective forces whose predictive power has already been shown 
when describing ground state nuclear properties all over the nuclear chart. 
In this context, the set of effective interactions already mentioned is well 
suited and it is very interesting to compare their predictions. In a second step, 
the role played by the $\gamma$ degree of freedom will also be discussed. The 
corresponding triaxial calculations, will be based on the code EV8 \cite{ev8}, 
our main computational tool in the present work. We will take advantage of its 
three dimensional lattice discretization that allows to treat any quadrupole 
deformation effect, axial or triaxial, on the same footing 
\cite{flocard,Bender-Review}. On the other hand, these calculations are much 
more involved than the  axial ones (about a factor 60 much more expensive in 
computing time) and therefore have been restricted to a relevant sample of 
nuclei in the present study. 

The paper can be outlined as follows. In Sec. \ref{T-FRAMEWOK} we present a 
brief description of the main theoretical formalism (Hartree-Fock + BCS) used 
to obtain the main ingredients of the present study, i.e., the PECs and 
potential energy surfaces (PESs) for the considered isotopic chains. For more 
details, the reader is referred to the corresponding literature. 
Sec. \ref{Results} contains our results. There, we will first discuss 
the sensitivity of the PECs (obtained in the framework of axially symmetric 
calculations) to the effective nucleon-nucleon force and to the treatment of 
the pairing correlations while in a second step the role of triaxiality 
(i.e., the $\gamma$ degree of freedom) will be illustrated for nuclei of
the Yb, Hf, W, Os, and Pt chains with neutron numbers $N$=114,116, and 118. 
Finally, Sec. \ref{Conclusions} is devoted to the concluding remarks and 
work perspectives.

\section{Theoretical framework}
\label{T-FRAMEWOK}
 
Our microscopic approach is based on a self-consistent formalism built on a 
deformed Hartree-Fock mean-field, using Skyrme type two-body interactions, 
plus pairing correlations between like nucleons included in the BCS 
approximation. As it is well known, the density-dependent HF+BCS approximation 
provides a very good description of ground-state properties for both spherical 
and deformed nuclei \cite{flocard} and it is at present one of the possible 
state of the art mean-field descriptions \cite{Bender-Review}.

We have used two methods of solving the deformed HF+BCS equations. The first 
method, is based on the expansion of the single-particle wave functions into 
an appropriate orthogonal basis (the eigenfunctions of an axially symmetric 
harmonic oscillator potential), following the procedure based on the formalism 
developed in Ref. \cite{vautherin}. The second  method, which is our main
choice for the calculations performed in this study, uses a coordinate space 
mesh, solving the HF+BCS equations for Skyrme type functionals via 
discretization of the individual wave functions on a 3-dimensional cartesian 
mesh \cite{ev8}. As a matter of fact, it can be shown that this corresponds
to an expansion on the basis of Lagrange polynomials associated with the
selected 3-dimensional mesh \cite{3D-ref}. One of the main advantages of the  
3-dimensional discretization is that any quadrupole deformation effect, axial 
or triaxial, can be taken into account on the same footing 
\cite{flocard,Bender-Review}. As a result, this second method represents our 
main computational tool in the present study and will be used later on, when 
examining the role played by the $\gamma$ degree of freedom for the considered
nuclei. 

We have mainly considered three different parametrizations of the effective 
Skyrme-like interactions in the particle-hole channel. As the leading choice, 
we have performed calculations (both axial and triaxial) with the parametrization 
SLy4 \cite{sly4}. We also show results in some instances (i.e., at the level of 
axially symmetric calculations) for the Skyrme forces Sk3 \cite{sk3} and SLy6
\cite{sly4}. They are examples of global effective Skyrme interactions that have 
been designed to fit ground state properties of spherical nuclei and nuclear 
matter properties. While Sk3 is the most simple one, involving in particular a 
linear dependence on the density, the  Lyon force SLy4  represents a 
parametrization obtained with a more recent fitting protocol and its
predictive power has already been shown to be very reasonable all over the 
nuclear chart \cite{Bender-Review}. Calculations have also been 
performed with the Skyrme parameter set SLy6, which includes additionally a 
two-body center of mass correction in the corresponding energy functional. 

As a first step, HF+BCS calculations preserving axial symmetry have been
performed using the Skyrme forces mentioned above. Our plan is, on the one hand, 
to use  the axially symmetric calculations to study the sensitivity of our 
results to the effective interactions used in the particle-hole and 
pairing channels. On the other hand, we will also use them to obtain, in the 
framework of our Skyrme HF+BCS study, first hints on shape transitions around 
$^{190}$W. Obviously, pairing correlations have also been taken into account 
and we have selected a zero-range density-dependent pairing force \cite{terasaki}, 

\begin{equation}  
V({\bf r_1},{\bf r_2})=-g \left( 1-\hat{P}^{\sigma} \right)
\left( 1-\frac{\rho({\bf r_1} )}{\rho_c}\right)
\delta ( {\bf r_1}- {\bf r_2}  )\, ,
\label{dd-pairing}   
\end{equation}
as our leading choice in the particle-particle channel. In Eq. (\ref{dd-pairing})
$\hat{P}^{\sigma}$ is the spin exchange operator, $\rho({\bf r})$ is the nuclear 
density, and $\rho_c=0.16$ fm$^{-3}$. The strength  $g$ of the pairing force 
[Eq. (\ref{dd-pairing})] is taken $g=1000$ MeV fm$^3$ for both neutrons and 
protons and a smooth cut-off of 5 MeV around the Fermi level has been introduced
\cite{terasaki,Rigo}. Let us mention that very recently the parametrization SLy4 
has been successfully applied in combination with the pairing interaction 
[Eq. (\ref{dd-pairing})]  (with $g=1000$ MeV fm$^3$) in systematic studies of 
correlation energies from $^{16}$O to the superheavies \cite{other1} and in 
global studies of spectroscopic properties of the first $2^{+}$ states in 
even-even nuclei \cite{other2}. Thus, the predictive power of this combination 
of effective interactions, has been well established along the nuclear chart, 
and this is the main reason for selecting the combination SLy4 in the 
particle-hole channel and the interaction of Eq. (\ref{dd-pairing}) (with $g=1000$ 
MeV fm$^3$) in the pairing channel as the leading choice for the present study. 
Calculations with the parametrization SLy6 in the particle-hole channel  
and the interaction of Eq. (\ref{dd-pairing}) (with $g=1000$ MeV fm$^3$)  have also been
performed.

In this work we also consider other recipes for pairing correlations. For example, 
a schematic seniority pairing force with a constant pairing strength $G$ 
parametrized to reproduce the phenomenological pairing gaps will also be  used. 
This treatment is called constant-force approach. One can also further simplify 
the pairing treatment by parametrizing the pairing gaps $\Delta_{p,n}$ directly 
from experiment, we call this treatment constant-gap approach. The pairing 
strength $G$ and the pairing gap are related through the gap equation \cite{rs}

\begin{equation}
\Delta = G \sum_{k >0} u_{k} v_{k} \, ,
\end{equation}
where $v_{k}$'s are the occupation amplitudes  ($u_k^2=1-v_k^2$).

The PECs shown later on in this paper (see Section \ref{Results}) are computed 
microscopically by constrained HF+BCS calculations \cite{rs,constraint} in which 
axial symmetry is kept as self-consistent symmetry \cite{rs} during the typical 
mean field iterative procedure \cite{Bender-Review}.  

In a second step, calculations exploring the role played by the $\gamma$ degree 
of freedom have been performed \cite{ev8} for the Yb, Hf, W, Os, and Pt chains 
with neutron numbers $N$=114,116, and 118. As mentioned before, the triaxial 
calculations are more involved than the axial ones and we have restricted them to 
that selected set of nuclei. As in the axially symmetric 
calculations, the corresponding energy functional is minimized under a quadratic 
constraint that holds the mass quadrupole moment fixed to a given value (expressed 
in barns) \cite{ev8,rs} and the non linear HF+BCS equations are solved using the 
method of successive iterations \cite{Bender-Review,ev8}. In addition to the usual 
mean field constraints on both neutron and proton numbers, the nuclear shape is 
determined by constraining simultaneously the pair of values $(q_{1},q_{2})$ related 
with the parameters $Q$ and $\gamma$ through the equations \cite{ev8}

\begin{eqnarray} \label{eq-q1q2}
q_{1} &=& Q \left( \cos \gamma - \frac{1}{\sqrt{3}}  \sin \gamma \right)
\nonumber\\
q_{2} &=& Q \frac{2}{\sqrt{3}}  \sin \gamma 
\nonumber\\
Q &=& \sqrt{q_{1}^{2} + q_{2}^{2} + q_{1}q_{2}}
\end{eqnarray}

Particular shapes of interest are the prolate ones with 
$\gamma = 0^{o}$ ($(q_{1}=Q,q_{2}=0)$ ) and the oblate ones with 
$\gamma = 60^{o}$ ($(q_{1}=0,q_{2}=Q)$ ), while triaxial shapes lie in between  
them. Using chains of triaxial calculations, we have checked the stability of 
the minima predicted in the framework of the axially symmetric calculations 
(first step in the present study) with respect to the $\gamma$ degree of freedom. 
These calculations will also be discussed in Sec. \ref{Results} and have been 
performed with the parametrization SLy4 plus the pairing interaction of Eq. 
(\ref{dd-pairing}) (with $g=1000$ MeV fm$^3$).

\section{Discussion of results}
\label{Results}

In this section we present a discussion of the results obtained in our study. 
First, we will discuss those obtained with the restriction to axially symmetric 
shapes and later on, in a second step, the role played by the $\gamma$ degree 
of freedom will also be illustrated.

%%%%%%%%%%%%%%%%%%%%%%%%%%%%Fig2%%%%%%%%%%%%%%%%%%%%%%%%%%%%%%%%%%%%%%%%%%%%%%%%%%%%%%%%%
\begin{figure}
\centering
\includegraphics[width=180mm]{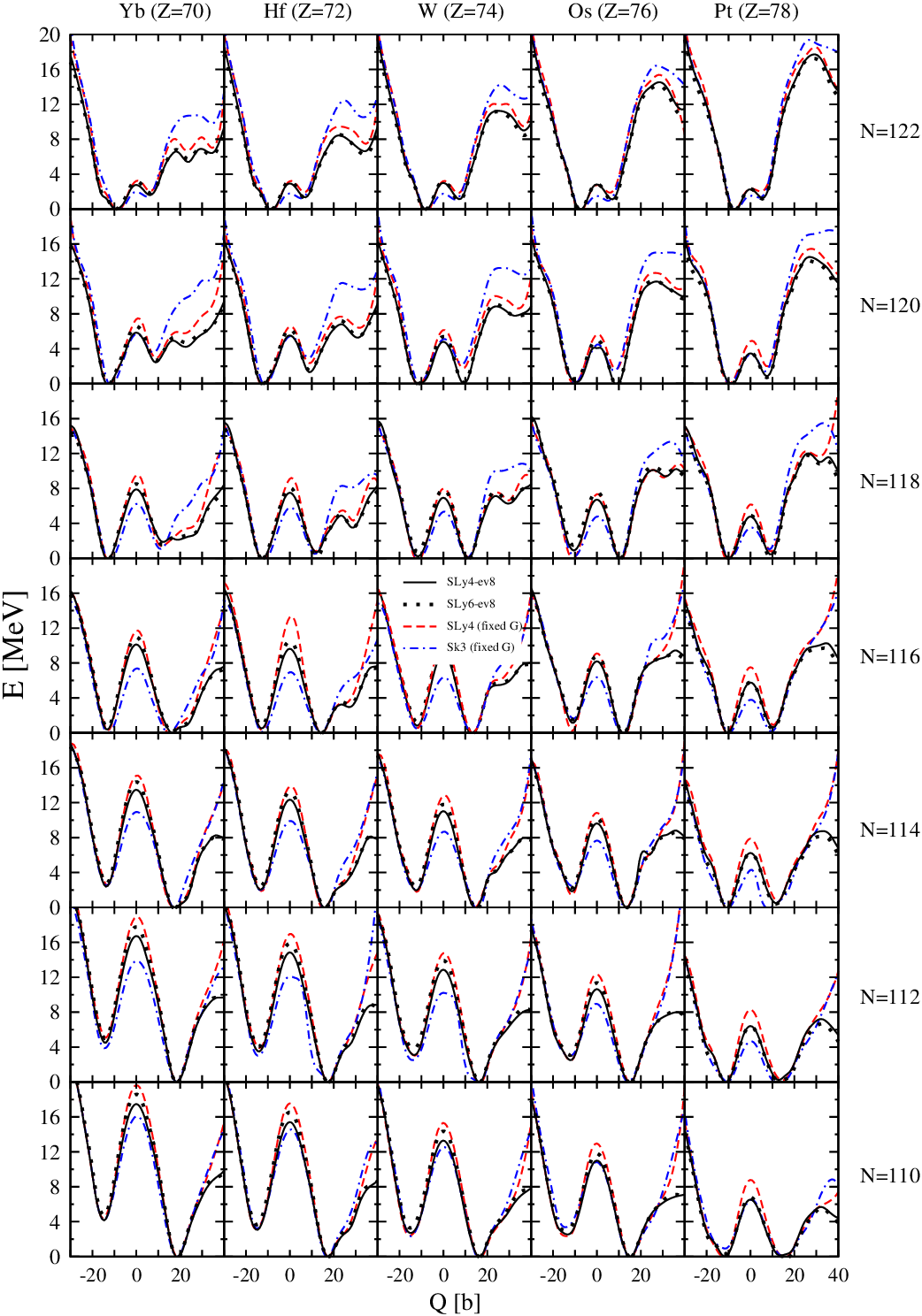}
\caption{(Color online) Potential energy curves for several isotope chains 
obtained from HF+BCS calculations with various Skyrme and pairing forces.
For details, see the main text in Section \ref{Results}. }
\label{figeq}
\end{figure}
%%%%%%%%%%%%%%%%%%%%%%%%%%%%%%%%%%%%%%%%%%%%%%%%%%%%%%%%%%%%%%%%%%%%%%%%%%%%%%%%%%%%%%%%%
%%%%%%%%%%%%%%%%%%%%%%%%%%%%Fig3%%%%%%%%%%%%%%%%%%%%%%%%%%%%%%%%%%%%%%%%%%%%%%%%%%%%%%%%%
\begin{figure}
\centering
\includegraphics[width=170mm]{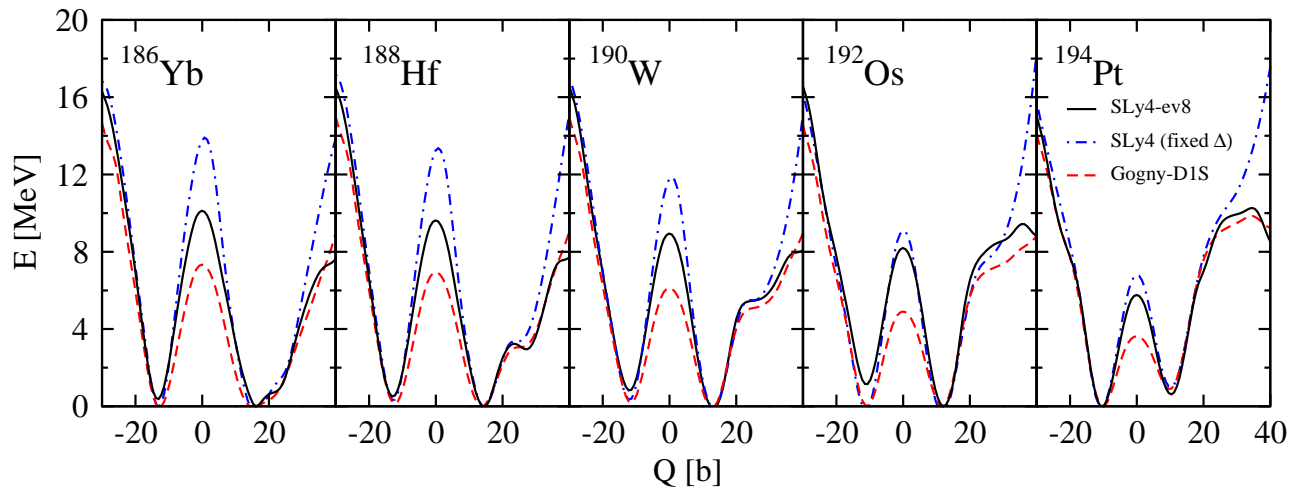}
\caption{(Color online) PECs for $N$=116 isotones obtained from mean field 
calculations with various forces. For details, see the main text  in 
Section \ref{Results}.}
\label{fign116}
\end{figure}
%%%%%%%%%%%%%%%%%%%%%%%%%%%%%%%%%%%%%%%%%%%%%%%%%%%%%%%%%%%%%%%%%%%%%%%%%%%%%%%%%%%%%%%%%

As it is well known, PECs are sensitive to the effective nuclear force in both 
relativistic \cite{meng} and nonrelativistic \cite{tajima,sarri} approaches, 
as well as to pairing correlations \cite{tajima,sarri,rrguzman}. Thus, we begin 
our discussion on the PECs by considering sample results in order to study the 
sensitivity of our predictions to effective interactions in both the particle-hole  
and the pairing channels. In Fig. \ref{figeq} we consider the PECs obtained in the 
framework of the axially symmetric calculations described in Sec. \ref{T-FRAMEWOK}
with the Skyrme forces Sk3, SLy4, and SLy6 and various pairing treatments. Taking 
as a reference the results obtained from the most sophisticated method based on 
the coordinate space mesh calculation \cite{ev8} with the forces SLy4 and SLy6 plus 
a zero-range density dependent pairing force with strength $g=1000$ MeV fm$^{3}$,
labeled SLy4-ev8 (solid lines) and SLy6-ev8 (dotted lines) in the figure, one 
can see that the results do not differ much when changing the pairing treatment 
to a constant-force approach (dashed lines) or when changing the Skyrme interaction 
into Sk3 (dash-dotted lines). It can also be seen, that the 
locations of the oblate and prolate minima appear at the same deformations, no 
matter what the force is. However, the relative energies can be slightly different 
and the spherical energy  barriers between the prolate and oblate minima of the 
PECs can change by a few MeV, depending on the force. In general, the energy 
barriers at zero deformation are lower with the force Sk3 and the delta pairing 
force makes also the barriers somewhat lower. These results indicate that the 
topology of the PECs is sensitive to the details of the calculations as has 
already been pointed out in Ref. \cite{rrguzman}. Additionally, looking at
 Fig. \ref{figeq}, it is clear that, at least for some of the nuclei considered, 
there is a very strong competition between different low-lying configurations 
corresponding to different intrinsic deformations (i.e., shape coexistence) and 
therefore dynamical correlations not explicitly taken into account at the mean 
field level (for example, symmetry restoration and/or configuration mixing)
could certainly  play a role in the description of ground state properties in 
these nuclei. 

If one analyzes Fig. \ref{figeq} in the vertical direction along isotope chains, 
one can see a clear evolution from prolate to oblate shapes for increasing  
neutron number. This transition takes place at $N$=116 in Yb and Hf isotopes, 
at $N$=116, and $N$=118 in W and Os isotopes, and it is a very soft transition in Pt 
isotopes. On the other hand, horizontally along isotone chains and exception 
made of the Pt isotopes, the isotones with $N$=122,120, and 118 are oblate, while 
those with $N$=114,112, and 110 are prolate. Isotones with $N$=116 are transitional 
nuclei with oblate and prolate minima at about the same energy and therefore
showing a more pronounced  shape coexistence. It is also worth noticing that 
the energy barriers at zero deformation decrease almost linearly with the number 
of protons as $Z$ increases in a given isotone chain. The same reduction is 
observed  in isotope chains as $N$ increases.

%%%%%%%%%%%%%%%%%%%%%%%%%%%%Fig4%%%%%%%%%%%%%%%%%%%%%%%%%%%%%%%%%%%%%%%%%%%%%%%%%%%%%%%%%
\begin{figure}
\centering
\includegraphics[width=180mm]{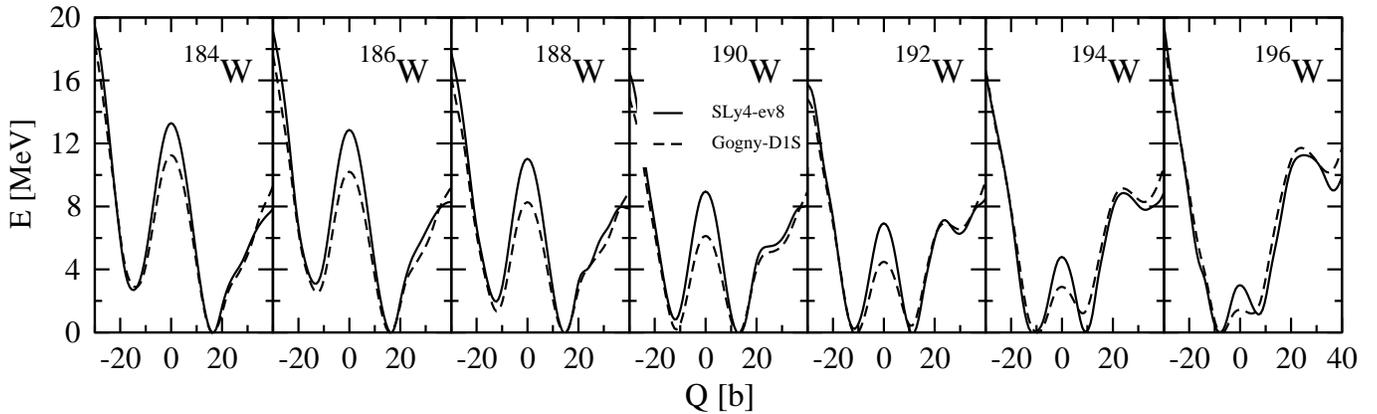}
\caption{PECs for $Z$=74 isotopes obtained from mean field calculations with various 
forces. For details, see the main text  in Section \ref{Results}.}
\label{figw}
\end{figure}
%%%%%%%%%%%%%%%%%%%%%%%%%%%%%%%%%%%%%%%%%%%%%%%%%%%%%%%%%%%%%%%%%%%%%%%%%%%%%%%%%%%%%%%%%

%%%%%%%%%%%%%%%%%%%%%%%%%%%%Fig5%%%%%%%%%%%%%%%%%%%%%%%%%%%%%%%%%%%%%%%%%%%%%%%%%%%%%%%%%
\begin{figure}
\centering
\includegraphics[width=110mm]{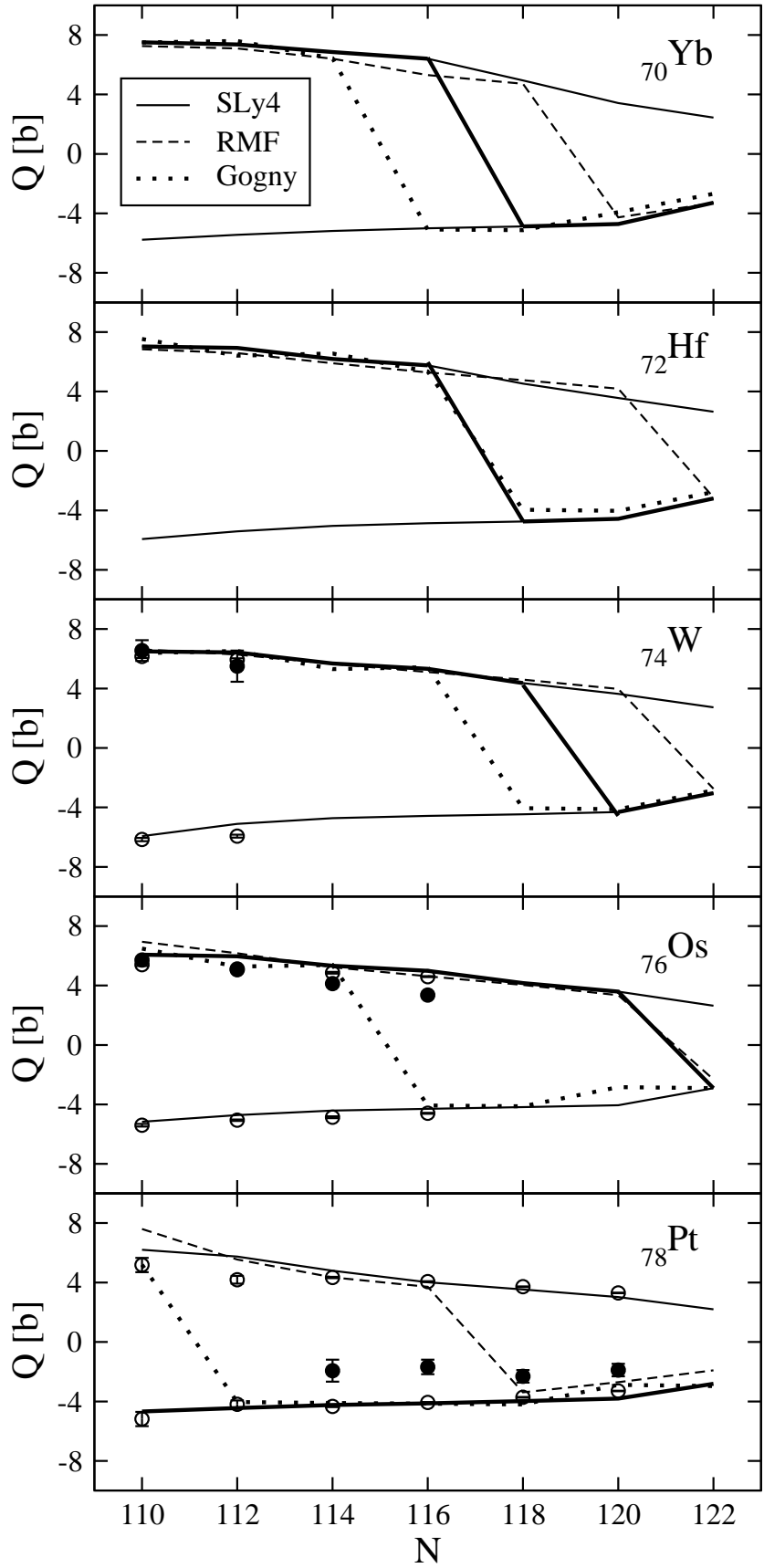}
\caption{Intrinsic quadrupole moments for five isotope chains as a function 
of the neutron number. Theoretical results obtained from Skyrme-SLy4 (solid lines), 
relativistic mean-field (dashed lines), and Gogny force (dotted lines) are compared 
to experimental data from Refs. \protect{\cite{stone}} (full circles) and
\protect{\cite{raman}} (open circles). For details, see text. In the case of 
SLy4, thick lines are used to connect the quadrupole moments of the ground states.}
\label{figqn}
\end{figure}
%%%%%%%%%%%%%%%%%%%%%%%%%%%%%%%%%%%%%%%%%%%%%%%%%%%%%%%%%%%%%%%%%%%%%%%%%%%%%%%%%%%%%%%%%

The previous  results are in qualitative good agreement with the ones obtained 
in Ref. \cite{fossion} using the parametrization NL3 of the RMF Lagrangian 
and a pairing force based on the Brink-Boeker part of the Gogny interaction. 
The energy barriers are found to be lower than SLy4, but the location of 
the equilibrium deformations does not change much. The agreement of our 
results with those in Ref. \cite{stevenson05} is also satisfactory concerning 
 the location of the minima, but the energy barriers at zero 
deformation are clearly lower in that reference. In general, our results 
are in good agreement with the ones obtained in previous studies. However, the
nuclides at which the transitions from one shape to another take place, may 
change because of the small energy differences between the oblate and prolate 
minima in the transitional region.

Figs. \ref{fign116} and \ref{figw} show a comparison of the PECs obtained from 
the force SLy4 and a zero-range density dependent pairing force (referred as 
SLy4-ev8) with those obtained from Hartree-Fock-Bogoliubov calculations 
\cite{egido-1,hilaire} based on the parametrization D1S of the Gogny interaction
\cite{d1s}. As it is well known, the advantage of the Gogny interaction over 
other alternatives, is that its finite range allows a fully self-consistent 
treatment of pairing correlations with the same interaction that produces, due 
to its structure, the proper cut off for the pairing matrix elements.

The comparison is done in Fig. \ref{fign116} for the isotone chain $N$=116 
while a similar comparison in done in Fig. \ref{figw} for the isotope chain 
$Z$=74. In Fig. \ref{fign116} we also show the results obtained with the Skyrme 
force SLy4 using a simple constant-gap approach in the pairing channel with 
pairing gaps obtained from the experimental masses of neighboring 
nuclei \cite{audi}. These figures show again that the 
location of the minima is rather stable at practically the same deformations.
It should also be noted that the energy barriers are somewhat lower with the 
Gogny interaction and somewhat larger with the constant-gap approach.

The quadrupole moments of the charge distributions of these nuclei can be 
compared with the available experimental information. This is done in 
Fig. \ref{figqn} for our five isotopic chains as a function of the neutron 
number. Our theoretical results obtained from SLy4 are shown by thin solid 
lines, one connecting the prolate minima and the other joining the oblate 
ones. Thick lines connect the quadrupole moments corresponding to the ground 
states of the nuclei. We also show by dashed lines the quadrupole moments of 
the ground states obtained from RMF calculations \cite{lala} (parametrization 
NL3) and the BCS formalism with constant pairing gaps. Dotted lines correspond 
to Gogny-D1S calculations \cite{hilaire}. 
Experimental data (full circles) are taken from the most updated compilation
of available data on static nuclear electric quadrupole moments \cite{stone}, 
transforming the laboratory quadrupole moments 
corresponding to the first $2^+$ excitations into intrinsic ones by using 
$Q_{\rm intr}=-3.5\; Q_{\rm lab}$. We also show the experimental intrinsic 
quadrupole moments derived from the experimental values of $B(E2)$ strengths 
\cite{raman}. In this case the sign cannot be extracted and therefore we show 
by open circles in Fig. \ref{figqn} both possibilities of signs. Actually, the 
$B(E2)$ strengths are converted into quadrupole moments \cite{raman} under the 
assumption of a well defined axial rotor behavior that in our case can be 
identified with values of the $E_{4^+}/E_{2^+}$ ratio close to 3.33.
It is worth noticing that in Ref. \cite{wu96} a complete set of $E2$ matrix
elements involving the low-lying excited states in $^{186,188,190,192}$Os 
and $^{194}$Pt was measured. This includes not only the $B(E2)$ values but also
the relative signs between the transitional $E2$ matrix elements and the
static quadrupole moments. In particular, it was shown that the $<Q^2>$ centroids 
are nearly spin-independent, suggesting that the $E2$ properties are correlated
and that the collective motion is rotation-like. The general trend of the data
in Ref. \cite{wu96} is consistent with the description in terms of $\gamma$-soft 
type collective models through a prolate to oblate shape transition. It would 
be very interesting in the future to include dynamical correlations beyond 
the static mean field picture used in this work and compare the results 
with those data.

As can be observed from Fig. \ref{figqn}, there is a nice agreement between our SLy4 
results and the experimental ones, including the sign (full circles). At the same 
time RMF calculations fail to describe the sign of $^{192,194}$Pt isotopes, 
while calculations with the Gogny interaction fail to describe the quadrupole 
moment of $^{192}$Os. Nevertheless, one should notice that in the transitional 
region this discrepancy is not very significant because the energies corresponding 
to the two deformations are practically the same and small details of the 
calculation may change the relative energy difference between the oblate and the 
prolate shapes. 

%%%%%%%%%%%%%%%%%%%%%%%%%%%%Table1%%%%%%%%%%%%%%%%%%%%%%%%%%%%%%%%%%%%%%%%%%%%%%%%%%%%%%%
\begin{table}[h]
\begin{center}
\caption{Ratio $E_{4^+}/E_{2^+}$ and moments of inertia  ${\cal I}$ [MeV $^{-1}$] 
obtained from the first
experimental $2^+$ energy (${\cal I}_{\rm exp}$), from a cranking calculation
with SLy4 (${\cal I}_{\rm SLy4}$) and Gogny (${\cal I}_{\rm Gogny}$), as well as 
from macroscopic descriptions of the nucleus as an irrotational flow 
(${\cal I}_{\rm i.f.}$) and a rigid rotor 
(${\cal I}_{\rm r.r.}$). Only nuclei with 
$E_{4^+}/E_{2^+} > 3$ are considered.}
\vskip 0.5cm
{\begin{tabular}{c|cccccc} 
\hline
\hline  & $E_{4^+}/E_{2^+}$ & ${\cal I}_{\rm exp}$ & ${\cal I}_{\rm SLy4}$ 
& ${\cal I}_{\rm Gogny}$ & ${\cal I}_{\rm i.f.}$ & ${\cal I}_{\rm r.r.}$  \\
\hline
$^{182}$Hf  & 3.29 & 30.7 & 31.8 & 31.5 & 5.0 & 87.9  \\
$^{184}$Hf  & 3.26 & 27.9 & 33.6 & 35.0 & 4.8 & 89.4  \\
$^{184}$W   & 3.27 & 27.0 & 30.0 & 27.6 & 4.3 & 89.0  \\
$^{186}$W   & 3.23 & 24.5 & 32.0 & 32.5 & 4.1 & 90.4  \\
$^{188}$W   & 3.09 & 21.0 & 33.0 & 23.3 & 3.2 & 91.3  \\
$^{186}$Os  & 3.16 & 21.9 & 29.2 & 26.8 & 3.8 & 90.2  \\
$^{188}$Os  & 3.08 & 19.4 & 31.6 & 23.5 & 3.6 & 91.6  \\
\hline
\end{tabular}}
\end{center}
\label{table1}
\end{table}
%%%%%%%%%%%%%%%%%%%%%%%%%%%%%%%%%%%%%%%%%%%%%%%%%%%%%%%%%%%%%%%%%%%%%%%%%%%%%%%%%%%%%%%%

Finally, we show in Table I the moments of inertia calculated microscopically 
within the cranking model, as well as the moments of inertia from various 
macroscopic models, namely, the rigid rotor (${\cal I}_{\rm r.r.}$) and the 
irrotational flow (${\cal I}_{\rm i.f.}$) models. The expressions used to 
calculate the moments of inertia can be found in Refs. \cite{rs,momin}:

\begin{eqnarray}
{\cal I}_{\rm cranking} & = & 2 \sum _{k,k'>0} \frac{\left| \langle k 
\left| J_x \right| k' \rangle \right| ^2 }{E_k+E_{k'}} \left( u_k v_{k'}
-u_{k'} v_k \right) ^2  \, ,\\
{\cal I}_{\rm r.r.} & = & \frac{2}{5}mAR_0^2\left( 1 + \sqrt{\frac{5}{16\pi}}
\beta \right) \, ,\\
{\cal I}_{\rm i.f.} & = & \frac{3}{5}\rho_0 R_0^5 \beta ^2 \, .
\end{eqnarray}
In these expressions $\beta$ is the quadrupole deformation defined in terms 
of the mass quadrupole moment $Q$ and the mean-square radius of the mass 
distribution $\langle r^2 \rangle$, 

\begin{equation}
\beta = \sqrt{\frac{\pi}{5}}\frac{Q}{A\langle r^2 \rangle }.
\end{equation}
We compare our results with the experimental moments of inertia extracted from 
the first $2^+$ excitations under the assumption that they correspond to 
rotors with $E_{2^+}=3/{\cal I}$  \cite{rs}. Thus, we quote only those nuclei 
with ratios $E_{4^+}/E_{2^+}$ close to the value 3.33. As expected, the rigid 
rotor and irrotational flow models predict, respectively, upper and 
lower limits to the phenomenological moment of inertia. The cranking moments 
of inertia, calculated either with SLy4 or Gogny forces, are much closer to 
experiment.

\begin{figure}
\centering
\includegraphics[width=110mm]{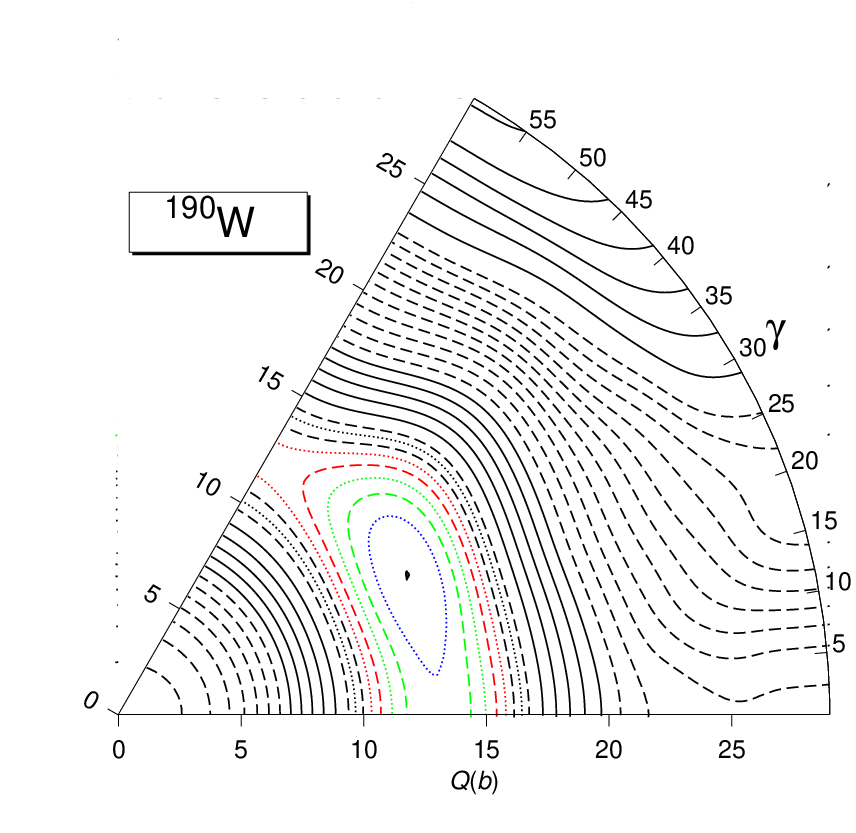}
\caption{(Color online) Contour plot of the potential energy surface of the 
nucleus $^{190}$W in a $\beta-\gamma$ plane-like representation. Instead of 
the  $\beta$ deformation parameter we have used the total mass quadrupole 
moment $Q$ (see Eq.(\ref{eq-q1q2}) in Sec. \ref{T-FRAMEWOK}) in units of barns. 
The contour lines correspond to the following scheme: from 
${\epsilon}_{min}+0.25$ MeV to ${\epsilon}_{min}+2.0$ MeV alternating dotted 
and dashed contours every  $0.25$ MeV (the contour ${\epsilon}_{min}+0.25$ MeV 
is plotted in blue, the contours ${\epsilon}_{min}+0.50$ MeV and 
${\epsilon}_{min}+0.75$ MeV are plotted in green, whereas those corresponding 
to ${\epsilon}_{min}+1.00$ MeV and ${\epsilon}_{min}+1.25$ MeV are plotted in red; 
the rest in black); from  ${\epsilon}_{min}+2.50$ MeV to ${\epsilon}_{min}+4.50$ MeV 
the contour lines are plotted every $0.50$ MeV as full lines; from 
${\epsilon}_{min}+5.00$ MeV to ${\epsilon}_{min}+10.00$ MeV the contour lines are 
plotted again every $0.50$ MeV  but this time as dashed lines; finally from
${\epsilon}_{min}+11.00$ MeV to ${\epsilon}_{min}+20.00$ MeV contour lines 
are depicted every $1.00$ MeV as full lines. Calculations have been performed 
with the parametrization SLy4 of the Skyrme force in the particle hole channel 
plus a zero range and density dependent pairing interaction with strength 
$g=1000$ MeV fm$^{3}$. For details, see the main text.
}
\label{beta-gamma-190W}
\end{figure}

\begin{figure}
\centering
\includegraphics[width=170mm]{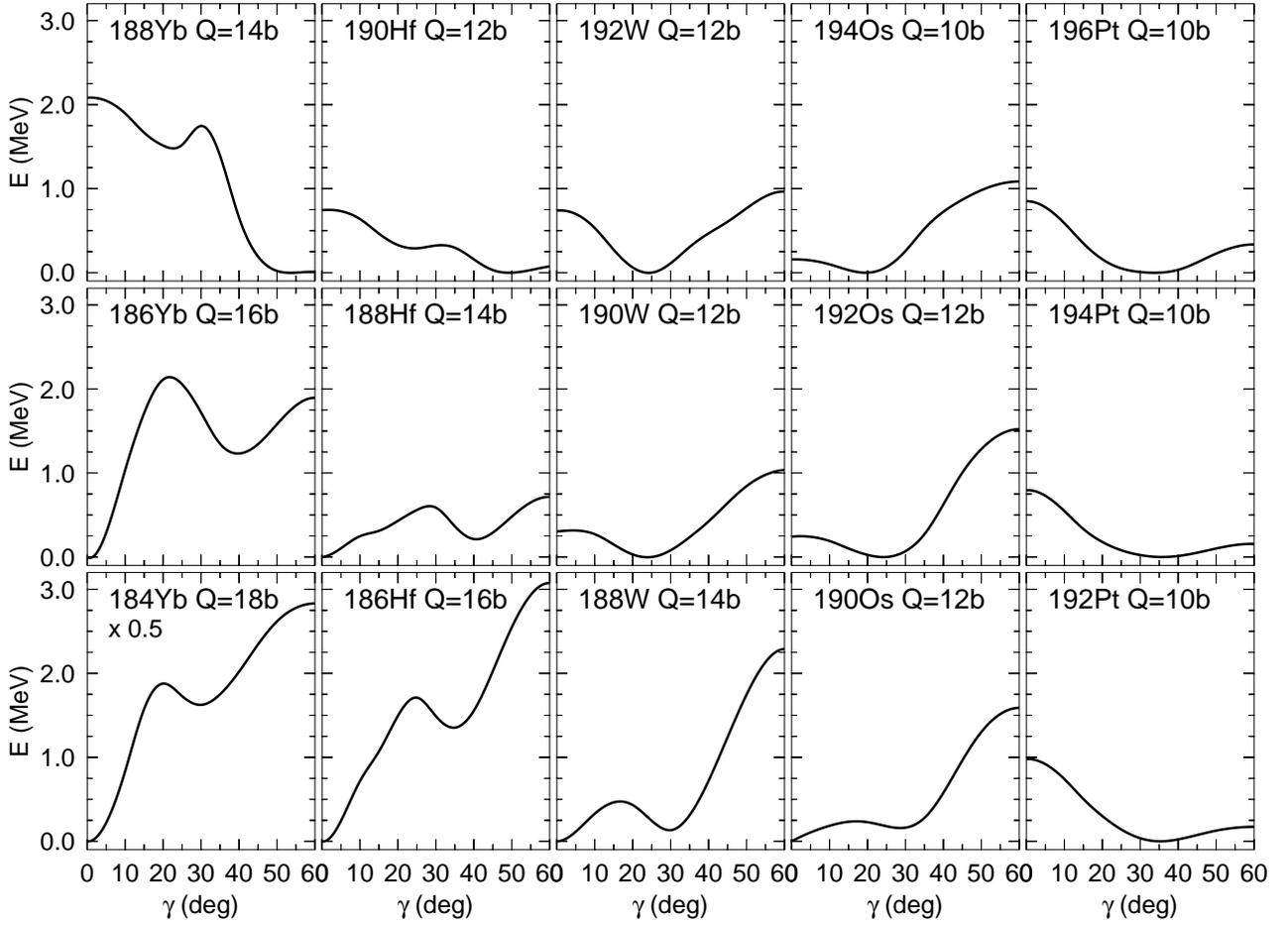}
\caption{HF+BCS energy of $N$=114, 116, and 118 isotones plotted as a function 
of the triaxial deformation parameter $\gamma$ (degrees) for the $Q$-values 
at which the energy minima are obtained in the axial case.}
\label{minima-q1q2}
\end{figure}

Let us turn now, to the second step of our discussion, i.e., the role played 
by the $\gamma$ degree of freedom in the considered nuclei. That triaxiality 
could certainly play a role for nuclei in this region of the nuclear chart, 
becomes already clear if one keeps in mind that the heavier $N$=116 isotones 
$^{192}$Os and $^{194}$Pt, are known to be $\gamma$-soft nuclei \cite{wu96}. 
The same applies to $^{190}$W from its $E_{4^+}/E_{2^+}$ ratio approaching 
the limit of 2.5 (see Fig. \ref{fige42}). Therefore, to confirm the 
reliability of the shape transitions predicted in the framework of the axially 
symmetric calculations already discussed above, we have carried out 
calculations constraining the $Q-\gamma$ degrees of freedom (instead of the 
$\beta$ deformation parameter we use the Q quadrupole moment) along the lines 
described in Sec. \ref{T-FRAMEWOK}. The calculations in 
Figs. \ref{beta-gamma-190W} and \ref{minima-q1q2} have been performed with  
the parametrization SLy4 in the particle hole channel plus the pairing 
interaction (Eq. (\ref{dd-pairing})) with $g=1000$ MeV fm$^{3}$.

%%%%%%%%%%%%%%%%%%%%%%%%%%%%Fig6%%%%%%%%%%%%%%%%%%%%%%%%%%%%%%%%%%%%%%%%%%%%%%%%%%%%%%%%%
%%%%%%%%%%%%%%%%%%%%%%%%%%%%%%%%%%%%%%%%%%%%%%%%%%%%%%%%%%%%%%%%%%%%%%%%%%%%%%%%%%%%%%%%%

%%%%%%%%%%%%%%%%%%%%%%%%%%%%Fig7%%%%%%%%%%%%%%%%%%%%%%%%%%%%%%%%%%%%%%%%%%%%%%%%%%%%%%%%%
%%%%%%%%%%%%%%%%%%%%%%%%%%%%%%%%%%%%%%%%%%%%%%%%%%%%%%%%%%%%%%%%%%%%%%%%%%%%%%%%%%%%%%%%%

In Fig. \ref{beta-gamma-190W} the contour plot for the PES of the nucleus 
$^{190}$W is presented. From this figure we immediately realize that the 
nucleus $^{190}$W has a triaxial ground state whose coordinates in the 
$Q-\gamma$ plane are $(Q,\gamma)=(14{\rm b}, 25^{o})$. The figure shows 
a soft behavior on the $\gamma$ degree of freedom with a very shallow 
triaxial minimum, which lies less than 0.5 (1.3) MeV below the prolate 
(oblate) minimum. To gain a better insight into the role of triaxiality 
in this mass region, we have performed similar calculations for other 
Yb, Hf, W, Os, and Pt nuclei with neutron numbers $N$=114,116, and 118. 
The results of these calculations are summarized in Fig. \ref{minima-q1q2}, 
where we plot the HF+BCS energy as a function of the triaxial deformation 
parameter $\gamma$ (in degrees) for the $Q$-values at which the energy 
minima are obtained in the axial case. In general, we can see that the 
oblate and prolate minima are connected in the $\gamma$ variable with 
smooth functions exhibiting in some cases shallow local minima or low 
peaks. The barriers found are of the order of a few hundred keV, much 
lower than the typical spherical barriers in the $\beta$ variable, which 
can reach values up to 20 MeV, depending on the example (see Fig. \ref{figeq}). 
In the case of Yb and Hf isotopes we can see a transition from prolate ($N$=114) 
to oblate ($N$=118) shapes, passing through a $\gamma$-soft isotope ($N$=116), 
which shows smooth peaks and valleys. We also observe that in these two
isotopes, one of the axial minima, oblate in $N$=114 and $N$=116 and prolate 
in $N$=118,
is indeed a saddle point, unstable in the $\gamma$-direction.
In the case of W isotopes we can see that the axial prolate and oblate minima
are again connected through soft curves in the $\gamma$ direction, 
allowing for triaxial minima, especially clear in the $N$=118 isotone. 
The oblate minima in $\beta$ become saddle points when the $\gamma$
degree of freedom is considered. The same happens for the $N$=116 and $N$=118
prolate minima.
Finally, Os (Pt) isotopes are examples of very soft prolate (oblate) nuclei 
with axial minima separated by about 1 MeV and connected through extremely 
shallow triaxial minima. 
Again, the oblate minima become saddle points and this is also the case for
the prolate ones, exception made of $^{190}$Os.
It is worth noticing once more that the axial minima are 
separated by spherical energy barriers ranging between 5 and 10 MeV.
The general trend seems to be quite clear: whenever there is a minimum,
which is much deeper than the other (2 MeV or more), the lower-lying 
minimum remains a minimum when the $\gamma$ degree of freedom is included.
On the other hand, the higher-lying minimum becomes a saddle point.
When the two minima lie at a depth not differing by more than a couple of
MeV, it is likely that a triaxial minimum develops and the prolate and 
oblate minima become saddle points. The latter is not a theorem but
represents a very likely situation that obviously is prone to exceptions.

\section{Conclusions}
\label{Conclusions}

With the aim to obtain first hints on nuclear phase shape transitions around 
$^{190}$W within the self-consistent Skyrme HF + BCS scheme, we have considered 
in the present work the PECs of five isotopic chains, namely, Yb, Hf, W, Os, and 
Pt for 106 $\le$ N $\le$ 122. Our study has been based on different Skyrme-like 
energy functionals plus different recipes for pairing correlations and comparisons
have also been undertaken with results obtained using the Hartree-Fock-Bogoliubov 
approximation based on the Gogny interaction.

From the analysis of our results we conclude that, at least in the mass region 
studied, the PECs are not sensitive to the method employed to solve the HF+BCS 
equations (3-dimensional cartesian lattice or deformed harmonic oscillator basis). 
We also conclude that the qualitative behavior of the energy profiles remains 
unchanged against changes in the Skyrme and pairing interactions in the sense 
that the deformations at which the energy minima occur are rather stable. This 
agrees well with results obtained with other nonrelativistic approximations 
including those involving the Gogny force, as well as with results obtained 
within the RMF. However, the spherical energy barriers between the minima are 
found to be sensitive to the details of the calculations.

As already mentioned, our main intention in this study has been to obtain 
first hints on nuclear phase shape transitions in the region under discussion. 
In this context, we find signatures for a  transition from prolate to oblate 
shapes as the number of neutrons increases from $N$=110 up to $N$=122 in Yb, 
Hf, W, and Os isotopes.
The lighter isotopes of these nuclei exhibit a rotational behavior that
changes gradually toward $\gamma$-soft as the number of neutrons increases.
The transition is found to happen at $N$=116-118, where the energies of 
prolate and oblate shapes are nearly degenerate. In the case of Pt, the 
isotopes considered do not show up a rotor behavior and a prolate shape in 
the lighter isotopes is not clearly developed yet. Let us also mention
that our results are in qualitative agreement with previous predictions on 
shape transitions in this mass region.

The role played by triaxiality in the description of ground state 
properties for nuclei in this mass region has been investigated 
by calculating $\beta-\gamma$ energy contour plots, and in particular,
the energy behavior with the $\gamma$ variable for fixed values of
$\beta$ corresponding to the axial minima. The analysis of our 
results shows that the axial prolate and oblate minima, which are
well separated by high spherical barriers in the $\beta$ degree of 
freedom, are linked very softly in the $\gamma$ degree of freedom.

Finally, there remains a long list of tasks to be undertaken in the 
near future and this work can be viewed as a starting point for a more 
ambitious project. In particular, it will be worth studying
up to what extent the results discussed above 
might be modified by including dynamical correlations beyond the static mean 
field picture. This is particularly interesting if one notes that, at least 
for some of the considered nuclei, shape coexistence is clearly visible and 
therefore it is important to check the role played by symmetry restoration 
and/or configuration mixing.  

\begin{acknowledgments}
This work was partly supported by Ministerio de Educaci\'on y Ciencia
(Spain) under Contract No.~FIS2005--00640 and 
Contract No.~FIS2004--06697. 
\end{acknowledgments}


\begin{thebibliography}{00}

\bibitem{review} J. L. Wood, K. Heyde, W. Nazarewicz, M. Huyse, and P. Van 
Duppen, Phys. Rep. {\bf 215}, 101 (1992).
\bibitem{gogny} J. Deharg\'e and D. Gogny, Phys. Rev. C {\bf 21}, 1568 (1980).
\bibitem{vautherin} D. Vautherin and D. M. Brink, Phys. Rev. C 
{\bf 5}, 626 (1972); D. Vautherin, Phys. Rev. C {\bf 7}, 296 (1973).
\bibitem{Bender-Review} M. Bender, P.-H. Heenen, and 
P.-G. Reinhard, Rev. Mod. Phys. {\bf 75}, 121 (2003).
\bibitem{lala-ref} D. Vetrenar, A. V. Afanasjev, G. A. Lalazissis, 
and P. Ring, Phys. Rep. {\bf{409}}, 101 (2005).  
\bibitem{gradient-egido} J. L. Egido, J. Lessing, V. Martin, and L. M. Robledo,
Nucl. Phys. A {\bf{594}}, 70 (1995).
\bibitem{podolyak} Zs. Podoly\'ak {\em et al.}, Phys. Lett. {\bf B491}, 
225 (2000).
\bibitem{caamano} M. Caama\~no {\em et al.}, Eur. Phys. J. A {\bf 23}, 201 (2005).
\bibitem{walker2006} P. M. Walker and F. R. Xu, Phys. Lett. {\bf B635}, 286 (2006).
\bibitem{nature} Ph. Walker and G. Dracoulis, Nature {\bf 399}, 35 (1999).
\bibitem{casten2000} R. F. Casten and B. M. Sherrill, Prog. Part. Nucl. Phys. 
{\bf 45}, S171 (2000).
\bibitem{jolie} J. Jolie and A. Linnemann, Phys. Rev. C {\bf 68}, 031301(R) (2003).
\bibitem{firestone} R. B. Firestone, C. M. Baglin, S. Y. Frank Chu, 
Table of Isotopes, 8th. ed., 1999 update (Wiley Interscience, 1999)
\bibitem{mach} H. Mach, Phys. Lett. {\bf B185}, 20 (1987).
\bibitem{wu96} C. Y. Wu {\em et al.}, Nucl. Phys {\bf A607}, 178 (1996).
\bibitem{naza90} W. Nazarewicz, M. A. Riley, and J. D. Garrett,  Nucl.
Phys. {\bf A512}, 61 (1990).
\bibitem{wheldon2000} C. Wheldon {\em et al.}, Phys. Rev. C {\bf 63}, 
011304(R) (2000).
\bibitem{naik} Z. Naik, B. K. Sharma, T. J. Jha, P. Arumugam, and S. K. Patra, 
Pramana {\bf 62}, 827 (2004).
\bibitem{fossion} R. Fossion, D. Bonatsos, and G. A. Lalazissis, 
Phys.Rev. C {\bf 73}, 044310 (2006).
\bibitem{stevenson05} P. D. Stevenson, M. P. Brine, Zs. Podolyak, P. H. Regan, 
P. M. Walker, and J. Rikovska Stone, Phys. Rev. C {\bf 72}, 047303 (2005).
\bibitem{stevenson01} P. Stevenson, M. R. Strayer, and J. Rikovska Stone, 
Phys. Rev. C {\bf 63}, 054309 (2001).
\bibitem{ev8} P. Bonche, H. Flocard, and P.-H. Heenen, Comput. Phys.
Comm. {\bf 171}, 49 (2005).
\bibitem{rs} P. Ring and P. Schuck, {\it {The Nuclear Many-Body Problem}} 
(Springer, Berlin-Heidelberg-New York) (1980).
\bibitem{sk3} M. Beiner, H. Flocard, N. Van Giai, and P. Quentin, 
Nucl. Phys. {\bf A238}, 29 (1975).
\bibitem{sly4} E. Chabanat, P. Bonche, P. Haensel, J. Meyer, and 
R. Schaeffer, Nucl. Phys. {\bf A635}, 231 (1998).
\bibitem{d1s} J. F. Berger, M. Girod, and D. Gogny,  Nucl. Phys. 
{\bf A428}, 23c (1984).
\bibitem{flocard} P. Bonche, H. Flocard, P. H. -Heenen, S. J. Krieger,
and M. S. Weiss, Nucl. Phys. {\bf A443}, 39 (1985). 
\bibitem{3D-ref} D. Baye and P.-H. Heenen, J. Phys. A {\bf{19}}, 2041 (1986).  
\bibitem{terasaki} J. Terasaki, P.-H. Heenen, H. Flocard, and P.
Bonche, Nucl. Phys. {\bf A600}, 371 (1996). 
\bibitem{Rigo} C. Rigollet, P. Bonche, H. Flocard, and 
  P.-H. Heenen, Phys. Rev. C {\bf 59}, 3120 (1999).
\bibitem{other1} M. Bender, G. F. Bertsch, and P.-H. Heenen, Phys. 
Rev. C {\bf 73}, 034322 (2006).  
\bibitem{other2} B. Sabbey, M. Bender, G. F. Bertsch, and P.-H. 
Heenen, Phys. Rev. C {\bf 75}, 044305 (2007).  
\bibitem{constraint} H. Flocard, P. Quentin, A. K. Kerman, and D.
 Vautherin, Nucl. Phys. {\bf A203}, 433 (1973).  
\bibitem{meng} J. Meng, W. Zhang, S. G. Zhou, H. Toki, and L. S. Geng,
Eur. Phys. J. A {\bf 25}, 23 (2005).
\bibitem{tajima} N. Tajima, S. Takahara, and N. Onishi, Nucl. Phys.
{\bf A603}, 23 (1996).
\bibitem{sarri} P. Sarriguren, O. Moreno, R. Alvarez-Rodriguez, and 
E. Moya de Guerra, Phys. Rev. C {\bf 72}, 054317 (2005).
\bibitem{rrguzman} R. Rodriguez-Guzman and P. Sarriguren, Phys. Rev. C {\bf 76},
064303 (2007).
\bibitem{egido-1} J. L. Egido, L. M. Robledo and 
R. Rodriguez-Guzman, Phys. Rev. Lett. {\bf 93}, 082502 (2004). 
\bibitem{hilaire} S. Hilaire, M. Girod, {\it Hartree-Fock-Bogoliubov results 
based on the Gogny force}, www-phynu.cea.fr
\bibitem{audi} G. Audi, O. Bersillon, J. Blachot, and A. H. Wapstra, Nucl. Phys.
{\bf A729}, 3 (2003).
\bibitem{lala} G. A. Lalazissis, S. Raman, and P. Ring, At. Data and
Nucl. Data Tables {\bf 71}, 1 (1999).
\bibitem{stone} P. Raghavan, At. Data Nucl. Data Tables {\bf 42}, 189 (1989); 
N. J. Stone, Table of Nuclear Moments (2001) 
${\rm www.nndc.bnl.gov/nndc/stone\_moments}$
\bibitem{raman} S. Raman, C. W. Nestor, Jr., and P. Tikkanen, At. Data Nucl. 
Data Tables {\bf 78}, 1 (2001).
\bibitem{momin} E. Moya de Guerra, Phys. Rep. {\bf 138}, 293 (1986).

\end{thebibliography}
\end{document}